\documentclass[final,3p,twocolumn,times,11pt]{elsarticle}
\usepackage{amssymb}
\usepackage{amsmath}
\usepackage{color}
\usepackage{xspace}

\journal{Physics Letters B}

\newcommand{\eqRef}[1]{eq.~(\ref{#1})\xspace}
\newcommand{\secRef}[1]{section~\ref{#1}\xspace}

\begin{document}

\begin{frontmatter}

\title{A framework for second-order parton showers }

\author[label1]{Hai Tao Li}
\ead{haitao.li@monash.edu}
\author[label1]{Peter Skands}
\address[label1]{ARC Centre of Excellence for Particle Physics at the Terascale, School of Physics and Astronomy, Monash University, VIC-3800 Australia}
\ead{peter.skands@monash.edu}

\begin{abstract}
A framework is presented for including second-order perturbative corrections to the radiation patterns of parton showers. The formalism allows to combine ${\cal O}(\alpha_s^2)$-corrected iterated  $2\to 3$ kernels for ``ordered'' gluon emissions with tree-level $2\to 4$ kernels for ``unordered'' ones.  The combined Sudakov evolution kernel is thus accurate to ${\cal O}(\alpha_s^2)$.  As a first step towards a full-fledged implementation of these ideas, we develop an explicit implementation of $2\to 4$ shower branchings in this letter.
\end{abstract}


\end{frontmatter}

% \linenumbers

\section{Introduction}
\label{sect:int}
Recent decades have seen tremendous improvements in our ability to combine fixed-order and resummed calculations in QCD. In the context of Monte Carlo event generators, the state of the art is now that several next-to-leading order (NLO) matrix elements can be combined with parton showers, with some progress even at the NNLO level. A new generation of shower algorithms has also been developed~\cite{Gustafson:1987rq,Lonnblad:1992tz,Sjostrand:2004ef,Nagy:2005aa,Nagy:2007ty,Giele:2007di,Dinsdale:2007mf,Schumann:2007mg,Winter:2007ye,Platzer:2009jq,Hoche:2015sya}, most of which are based on colour dipoles or antennae, however the formal accuracy of these showers remains governed by leading-order splitting functions~\cite{Jadach:2011cr,Hartgring:2013jma,Nagy:2014mqa,Jadach:2016zgk}. 

One option for improving the accuracy of the resummed part of the calculation is to match the shower evolution to higher-order analytical 
resummation~\cite{Alioli:2012fc}. 
In this letter, we instead propose to include higher-order corrections directly in the shower evolution, truncated in such a way that the full evolution kernels are accurate to ${\cal O}(\alpha_s^2)$. We construct a framework to include NLO corrections into the Sudakov form factor for final-state showers, which implies that more subleading logarithmic terms will be resummed. 
%This can be achieved as a combination of virtual corrections to the $2\to 3$ splitting functions and real corrections at the $2\to 4$ level. 
We do this by writing the Sudakov factor as a product of $2\to 3$ splittings, which are responsible for the ordinary strongly-ordered shower evolution (with one-loop corrections and tree-level $2\to 4$ corrections applied to products of $2\to 3$ branchings), and direct $2\to4$ ones which access parts of phase space that would look ``unordered'' from the iterated $2\to 3$ perspective and hence would not be reached by such branchings. 
A main stumbling block to achieving this in the past, which we address in this letter, is how to avoid double-counting between iterated $2\to 3$ branchings and direct $2\to 4$ ones. Our solution to this problem is to order the two branching types in a common measure of $p_\perp$, allowing a clean phase-space separation between them. We also ensure that there is a smooth transition at the boundary between the ordered and unordered regions. 

We work in the framework of dipole-antenna showers which combine dipole showers~\cite{Gustafson:1987rq} with the antenna subtraction formalism~\cite{Azimov:1986sf,Kosower:1997zr,GehrmannDeRidder:2005cm}, as embodied in the \textsc{Vincia} shower code\footnote{http://vincia.hepforge.org/}~\cite{Fischer:2016vfv,Sjostrand:2014zea}. \textsc{Vincia} was initially developed for final-state showers~\cite{Giele:2007di,Giele:2011cb,LopezVillarejo:2011ap,Hartgring:2013jma} and was recently extended to hadronic collisions~\cite{Ritzmann:2012ca,Fischer:2016vfv}. The aim of this letter is to demonstrate the basic formalism for second-order shower kernels (at leading colour) and provide a concrete proof-of-concept implementation of $2\to 4$ showers with two-gluon emission. We leave implementations of $g\to q\bar{q}$ splittings, one-loop corrections to $2\to 3$ showers, and a discussion of initial-state antennae to forthcoming work.

This letter is organised  as follows. In Section~\ref{sect:shower} we discuss the Sudakov factor and partition it into a product of $2\to 3$ and $2\to 4$ ones. Section~\ref{sect:2to4} presents the method for implementing $2\to 4$ branchings using the veto algorithm. In Section~\ref{sect:antenna} we describe the $2\to4$ antenna functions and compare them with corresponding matrix elements. In Section~\ref{sect:numeric} we discuss numerical results and collect our conclusions in Section~\ref{sect:conclu}~.

\section{Shower Framework} \label{sect:shower}
Within the existing antenna-shower formalism for a shower evolved in a generic measure of jet resolution $Q$, the LO subtraction term (antenna function) corresponding to a specific colour-connected pair of partons, call it  $a_3^0$~\cite{GehrmannDeRidder:2005cm}, is exponentiated to define an all-orders Sudakov factor, $\Delta(Q_1^2,Q_2^2)$, which represents the no-branching probability for that parton pair between scales $Q_1$ and $Q_2$. As such, the differential branching probability per phase-space element is given by the derivative of the Sudakov factor,
\begin{multline}\label{eq:sud}
\frac{d}{dQ^2}\left(1 - \Delta(Q_0^2,Q^2)\right) = \\-\int \frac{d\Phi_3}{d\Phi_2}  \ \delta(Q^2-Q^2(\Phi_3)) \ a_3^0 \ \Delta(Q_0^2,Q^2)~,
\end{multline}
where the $\delta$ function projects out a contour of constant $Q^2$ in the $2\to 3$ antenna phase space and we leave colour and coupling factors implicit in $a_3^0$. Typically, the phase space is then rewritten explicitly in terms of $Q$ and two complementary phase-space variables, which we denote $\zeta$ and $\phi$:
\begin{multline}
\frac{d\ln\Delta(Q_0^2,Q^2)}{dQ^2} = \int_{\zeta_{-}(Q)}^{\zeta_+(Q)} \!\!\!d\zeta \int_0^{2\pi} \frac{d\phi}{2\pi} \ \frac{|J| \ a_3^0}{16\pi^2 m^2}~,\label{eq:sudShower}
\end{multline}
with $m$ the invariant mass of the mother (2-parton) antenna. 
The Jacobian factor $|J|$ arises from the transformation to the $(Q,\zeta)$ variables and the $\zeta_{\pm}$ phase-space boundaries are defined by the specific choice of $Q$ and $\zeta$, see e.g.~\cite{Hartgring:2013jma}. It is now straightforward to apply more derivatives, in $\zeta$ and $\phi$, to obtain the fully differential branching probability in terms of the shower variables. 

The essential point is that, for $a_3^0$ to be the proper subtraction term for NLO calculations, it must contain all relevant poles corresponding to single-unresolved limits of QCD matrix elements. Thus, a shower based on $a_3^0$ is guaranteed to produce the same LL structure as DGLAP ones in the collinear limit~\cite{Nagy:2009re,Skands:2009tb}, while simultaneously respecting the dipole coherence embodied by the eikonal formula in the soft limit; the latter \emph{without} a need to average over azimuthal angles (as required for the angular-ordered approach to coherence, see e.g.~\cite{Ellis:1991qj}).

Generalising this formalism to use NNLO subtraction terms requires the introduction of the one-loop correction to $a_3^0$, call it $a_3^1$, as well as the tree-level double-emission antenna function, $a_4^0$. Explicit forms for all second-order antennae in QCD can be found in~\cite{GehrmannDeRidder:2005cm}, including their pole structure and factorisation properties in all single- and double-unresolved limits\footnote{Note that, for the 4-parton antenna functions, \cite{GehrmannDeRidder:2005cm} only provides  explicit formulae summed over permutations of identical gluons. These must then subsequently be partitioned into individual (sub-antenna) contributions from each permutation separately.}. Note that $a_3^1$ contains explicit singularities which appear as poles in $\epsilon$ in dimensional regularisation. These are cancelled by the poles in $a_4^0$ upon integration of one unresolved parton (while logarithms beyond those generated at LL will in general remain).

By analogy with \eqRef{eq:sud}, we define the differential branching probability as 
\begin{multline}
    \frac{d}{dQ^2} \Delta(Q_0^2,Q^2) = \\\int \frac{d\Phi_3}{d\Phi_2}
    \ \delta(Q^2-Q^2(\Phi_3)) \left(a_3^0 + a_3^1 - a_3^0\Delta^1 \right)   \Delta(Q_0^2,Q^2)\\
    + \int \frac{d\Phi_4}{d\Phi_2} \ \delta(Q^2-Q^2(\Phi_4))\ a_4^0 \  \Delta(Q_0^2,Q^2)~,\label{eq:sudNLO}
\end{multline}
where $\Delta^1$ is the ${\cal O}(\alpha_s)$ term in the expansion of
$\Delta(Q_0^2,Q^2)$, which would otherwise be double-counted, and $Q^2(\Phi_4)$ denotes the hardest clustering scale in $\Phi_4$, with the softer one being integrated over. Specifically, for a double clustering of $4 \to 3 \to 2$ partons, we define $Q(\Phi_4) \equiv \mathrm{max}(Q_4,Q_3)$; for an ordinary strongly ordered history, it is thus equal to the resolution scale of the clustered 3-parton configuration, $Q_3$, while for an unordered sequence, it is the 4-parton resolution scale, $Q_4$. 

We now come to the central part of our proposal: how to re-organise eq.~(\ref{eq:sudNLO}) in terms of finite branching probabilities (as mentioned above, the $a_3^1$ term and the integral over $a_4^0$ are separately divergent), expressed in shower variables and allowing iterated $2\to 3$ splittings and direct $2\to 4$ ones to coexist with the correct limiting behaviours (and no double counting) for both single- and double-unresolved emissions.

We first partition the $a_4^0$ function into two terms, one for each of the possible iterated $2\to 3$ histories, which we label $a$ and $b$ respectively. Suppressing the zero superscripts to avoid clutter, we define a $2\to 4$ correction factor in close analogy with the matrix-element-correction factors defined in~\cite{Giele:2011cb}, 
\begin{equation}
  R_{2\to 4} = \frac{a_4}{a_3 a'_3 + b_3 b'_3}~,\label{eq:R2to4}
\end{equation}
where $a_3$ and $b_3$ ($a'_3$ and $b'_3$) denote the antenna functions for the first (second) $2\to3$ splittings in the $a$ and $b$ histories, respectively. E.g., for 
$1_q 2_{\bar{q}} \to  3_q 4_g  5_g  6_{\bar{q}}$, 
the $a$ history is produced by the product of $a_3'(3,4,5)$ and $a_3(\widehat{34},\widehat{45},6)$, with the (on-shell) momenta of the intermediate 3-parton state, $\widehat{34}$ and $\widehat{45}$, defined by the phase-space map of the shower / clustering algorithm. 
The $b$ history is produced by the product of $b'_3(4,5,6)$ and $b_3(3,\widehat{45},\widehat{56})$. We emphasise that the denominator of eq.~(\ref{eq:R2to4}) is nothing but the incoherent sum of the $a$ and $b$ antenna patterns (modulo the ordering variable), as would be obtained from the uncorrected (LL) antenna shower, while the numerator is the full (coherent) $2\to 4$ radiation pattern. Among other things, the factor $R_{2\to 4}$ therefore contains precisely the modulations that account for coherence between colour-neighbouring antennae.

We use the definition of $R_{2\to 4}$, eq.~(\ref{eq:R2to4}), to partition $a_4$ into two terms, $a_4 = R_{2\to 4}~(a_3 a'_3 + b_3 b'_3)$, each of which isolates a specific (colour-ordered) single-unresolved limit, corresponding to either $g_4$ or $g_5$ becoming soft, respectively. For each term we iterate the exact antenna phase-space factorisation~\cite{GehrmannDeRidder:2005cm},
\begin{multline}
d\Phi_{m+1}(p_1,\ldots,p_{m+1}) = \\
d\Phi_m(p_1,\ldots,p_{I},p_{K},\ldots,p_{m+1}) \times d\Phi_\mathrm{ant}(i,j,k)~,
\end{multline}
with all momenta on shell and $p_i+p_j+p_k = p_I+p_K$, to write 
\begin{multline}
\frac{d\Phi_4(3,4,5,6)}{d\Phi_2(1,2)} = \\
\left\{ \begin{array}{ll}
 \mbox{path a:} & d\Phi_\mathrm{ant}(\widehat{34},\widehat{45},6) \ d\Phi_\mathrm{ant}(3,4,5) \\
\mbox{path b:} & d\Phi_\mathrm{ant}(3,\widehat{45},\widehat{56}) \ d\Phi_\mathrm{ant}(4,5,6) \end{array}
\right.~,\label{eq:PS2to4}
\end{multline}
where we have chosen the nesting of the antenna phase spaces 
such that the soft parton in the given history is always the one clustered first. We also divide up each of the resulting 4-parton integrals into ordered and unordered clustering sequences, for which $Q(\Phi_4)=Q_3$ and $Q(\Phi_4)=Q_4$, respectively (see above).
The result is
\begin{multline}
\label{eq:nlo_su}
 \frac{d\Delta(Q_0^2,Q^2) }{dQ^2}= \int d\Phi_\mathrm{ant}~\Bigg[
   \delta(Q^2 - Q^2(\Phi_3)) \ \Bigg( a_3^0 \\
  +a_3^1 - a_3^0\Delta^1
   + a_3^0\sum_{s\in a,b}\int_\mathrm{ord} \!\!\! d\Phi_\mathrm{ant}^s \ R_{2\to 4} \ s_3' ~\Bigg)\ \Delta(Q_0^2,Q^2)\\
   + \sum_{s\in a,b}\int_\mathrm{unord} \!\!\!\!\!\! d\Phi_\mathrm{ant}^s \delta(Q^2 - Q^2(\Phi_4)) R_{2\to 4} s_3 s_3' \Delta(Q_0^2,Q^2)\Bigg]
\end{multline}
where the sums in the last two lines run over the clustering sectors (= histories), $a$ and $b$.

We may now interpret the first two lines as an effective second-order probability density for $2\to 3$ branchings, while the last line represents a contribution from direct $2\to 4$ branchings.  The solution of eq.~(\ref{eq:nlo_su}) can be written as the product of $2\to 3$ and $2\to 4$ Sudakov form factors 
\begin{align}
   \Delta(Q_0^2,Q^2) = \Delta_{2\to 3}(Q_0^2,Q^2) \Delta_{2\to 4}(Q_0^2,Q^2)~.
\end{align}
Using the same notation as in \eqRef{eq:sudShower} and 
with $Q_3$ denoting a 3-parton resolution scale, the second-order $2\to 3$ Sudakov factor is:
\begin{multline}
\label{eq:2to3sudakov}
  \Delta_{2\to 3}(Q_0^2,Q^2) = 
  \exp \Bigg[
      -\int_{Q^2}^{Q^2_0}\!dQ^2_3 \int_{\zeta_-(Q_3)}^{\zeta_+(Q_3)} 
     \!d\zeta\
     \\
      \times
     \frac{|J|}{16\pi^2 m^2}  \bigg( a_3^0 + a_3^1
      + a_3^0 \sum_{s\in a,b}\int_\mathrm{ord} \!\!\! d\Phi_\mathrm{ant}^s \ R_{2\to 4} \ s_3'
      \\
      + a_3^0\int_{Q_3^2}^{Q_0^2} \!d\tilde{Q}_3^{2} \int_{\zeta_-(\tilde{Q}_3)}^{\zeta_+(\tilde{Q}_3)} \! d\tilde{\zeta} \ \frac{|\tilde{J}|}{16\pi^2 m^2}\ a_{\tilde{3}}^0 
      \bigg)
  \Bigg]~,
\end{multline}
where we have now written out the explicit form of the
$\Delta^1(Q_0^2,Q^2)$ term in the last line, with $|\tilde{J}|\equiv |J(\tilde{Q},\tilde{\zeta})|$. The functional form of $\tilde{Q}$ must be the same as that of $Q$ while the form of $\tilde{\zeta}$ can in principle be chosen independently of that of $\zeta$.  

The $2\to 4$ Sudakov factor is defined by the last term in \eqRef{eq:nlo_su}. However since the $\delta(Q^2 - Q^2(\Phi_4))$ function projects out the 4-parton resolution scale in this case, we interchange the order of the nested phase-space integrations, utilising that
\begin{multline}
      \int_0^{Q_0^2} \!dQ_3^2 \int_{Q^2}^{Q_0^2} \!dQ_4^2 \ \Theta(Q_4^2 - Q_3^2) ~ f(Q_3^2,Q_4^2)~ =\\ 
      \int_{Q^2}^{Q_0^2} \! dQ_4^2 \int_{0}^{Q_4^2} \! dQ_3^2~f(Q_3^2,Q_4^2)~,
\end{multline}
for a generic integrand, $f$, with the result:
\begin{multline}
   \Delta_{2\to 4}(Q_0^2,Q^2) = \exp     \bigg[ -\!\sum_{s\in a,b} \int_{Q^2}^{Q_0^2}\!dQ_4^2 \int_{0}^{Q_4^2} \!dQ_3^2 
   \\
   \int_{\zeta_{4-}}^{\zeta_{4+}} \!\!\!d\zeta_4 \int_{\zeta_{3-}}^{\zeta_{3+}}\!\!\!d\zeta_3 \ \frac{|J_3 J_4|}{(16\pi^2)^2m^2 m_s^2}\int_0^{2\pi}  \!\frac{d\phi_4}{2\pi} R_{2\to 4} s_3 s_3'
    \bigg]~,\label{eq:sud24}
\end{multline}
where the nested antenna phase spaces of \eqRef{eq:nlo_su}, $d\Phi_\mathrm{ant}\ d\Phi_\mathrm{ant}^s$ have now been expressed in terms of shower variables, with an associated combined Jacobian $|J_3J_4|$. 
In \secRef{sect:2to4}, we show how to construct an explicit shower algorithm based on \eqRef{eq:sud24} while we refer to \cite{Hartgring:2013jma} for a proof of concept of an NLO-corrected $2\to 3$ shower based on a formula that only differs from \eqRef{eq:2to3sudakov} by finite terms.

Let us now turn our attention to whether the integrands in each of the Sudakov form factors, eqs.~(\ref{eq:2to3sudakov}) and (\ref{eq:sud24}), are well-defined and finite. For $\Delta_{2\to 3}$, this amounts to showing whether the singularities present in the $a_3^1$ term are fully cancelled by those coming from the integral over $R_{2\to 4}s_3'$. We start from the observation that the single-unresolved limits of the 4-parton antenna functions are fully captured by the LL $2\to 3$ ones (up to angular terms which cancel upon integration over the unresolved region~\cite{GehrmannDeRidder:2005cm}), hence 
\begin{equation}
 a_4  ~\to~ a_3 a_3' ~+~ b_3 b_3' ~+~ \mathrm{ang.} ~,
\end{equation}
which in turn implies that $R_{2\to 4} \to 1$ in any single-unresolved limit (modulo the angular terms), hence the pole structure of the $R_{2\to 4}s_3'$ integrals is the same as that of the unmodified antenna functions, 
\begin{equation}
     \mathrm{Poles}\left\{ \int_\mathrm{ord} \!\!\!\! d\Phi_\mathrm{ant}^s \ R_{2\to 4} \ s_3' \right\} 
    = \mathrm{Poles} \left\{ \int\! d\Phi_\mathrm{ant}^s \ s_3' \right\},\label{eq:poles}
\end{equation}
where the integration region can be extended to all of phase space since the ordered region by definition includes all single-unresolved limits\footnote{This is true for all evolution variables considered in \textsc{Vincia} and, more generally, for any evolution variable that defines an infrared safe observable. Without this property, an explicit regularisation has to be introduced, see~e.g., the case of energy ordering considered in \cite{Hartgring:2013jma}.}, and use of the angular-averaged $R_{2\to 4}$ is justified since $s_3'$ itself does not depend on the azimuth angle. The sum of two sub-antenna integrals like the ones on the right-hand side of eq.~(\ref{eq:poles}) precisely cancels the singularities of the corresponding one-loop antenna functions, $a_3^1$~\cite{GehrmannDeRidder:2005cm}, thus establishing that the integrand in eq.~(\ref{eq:2to3sudakov}) is free of poles in $\epsilon$.
 
In the unordered part of phase space, singularities only occur when both $Q_4\to 0 \ \& \  Q_3 \to 0 $ which corresponds to part of  the double-unresolved contribution. In the shower context, these singularities are controlled via the assumption of unitarity. Thus, the $2\to4$ Sudakov factor is also well defined. Since the NLO $2\to3$ and $2\to4$ contributions are therefore both free of explicit poles in $\epsilon$, and since they generate corrections in different parts of phase space, they may be developed as separate algorithms, provided they use the same set of antenna functions. (Full second-order precision is of course only achieved when both components are included.) Given that a proof-of-concept study of NLO corrections to $\Delta_{2\to 3}$ already exists~\cite{Hartgring:2013jma}, we focus in the following sections on the previously missing piece: explicit construction of the $2\to 4$ component. 

\begin{figure}
  \centering
  \includegraphics[scale=0.5]{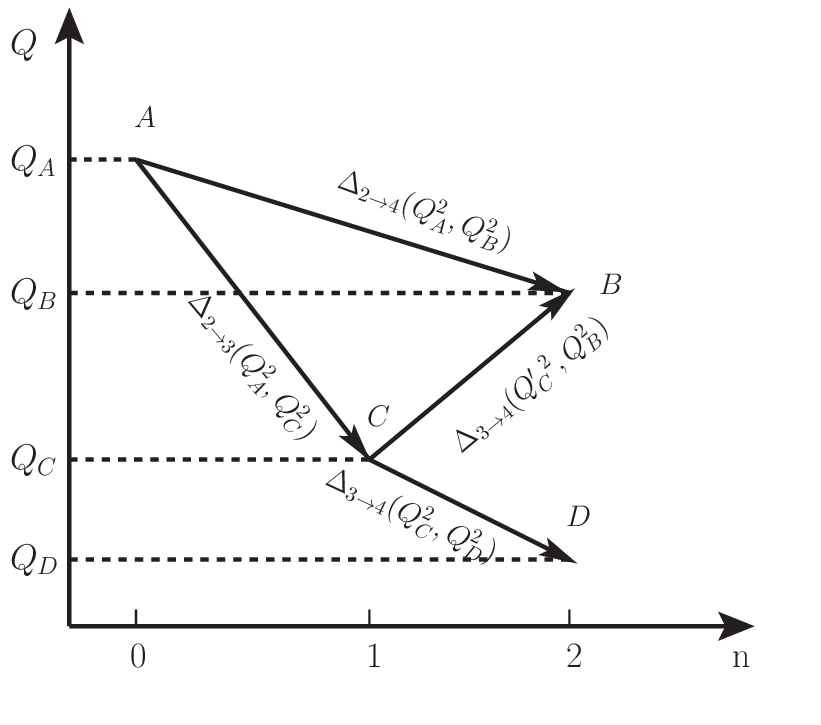} 
  \vspace{-2ex}
  \caption{\label{fig:scales}
  Illustration of scales and Sudakov factors in strongly ordered (ACD), smoothly (un)ordered (ACB), and direct $2\to 4$ (AB) branching processes, as a function of the number of emitted partons, $n$. 
  }
\end{figure}

We round off the discussion of the Sudakov form factors by illustrating the scale evolutions for $2\to 3$ and $2\to 4$ showers in fig.~\ref{fig:scales}. An ordered sequence of $2\to 3$ branchings is represented by path $A\to C\to D$ and the corresponding combined Sudakov factor is $\Delta_{2\to 3}(Q_A^2,Q_C^2) \Delta_{3\to 4}(Q_C^2,Q_D^2)$~. The $2\to 4$ shower explores more phase space by including path $A\to B$~ which lives in unordered phase space compared with the ordinary strongly-ordered shower. Path $A\to C\to B$  shows the possible branching in ``smoothly-ordered showers''~\cite{Giele:2011cb} which can also access unordered phase space. However, for smooth ordering the combined Sudakov factor $\Delta_{2\to 3}(Q_A^2,Q_C^2) \Delta_{3\to 4}(Q_C^{\prime~2},Q_B^2)$ is used  where $Q_C^{\prime} > Q_B$ represents the restart scale of the smooth-ordering shower. As pointed out in \cite{Hartgring:2013jma}, the $\Delta_{2\to 3}(Q_A^2,Q_C^2)$ factor implies an LL sensitivity to the intermediate scale $Q_C$; an undesired byproduct of the use of iterated on-shell $2\to 3$ phase-space factorisations. The direct $2\to4$ shower avoids this by using the exact Sudakov factor $\Delta_{2\to 4}(Q_A^2,Q_B^2)$ in which $Q_C$ only appears implicitly as an auxiliary integration variable. 

Finally, let us consider what happens in the vicinity of the boundary between what we label as ordered and unordered emissions, i.e., when there is no ``strong'' ordering between two successive (colour-connected) emissions. This is particularly relevant for the double-unresolved limits characterised by a single unresolved scale. The boundary can be approached either from the unordered region, or from the ordered one, and in general both regions will contribute to the double-unresolved limits. In the unordered region, the $2\to 4$ antenna functions are used directly, capturing both the single- and double-unresolved (soft and collinear) limits of QCD~\cite{GehrmannDeRidder:2005cm}. They are also in our formalism intrinsically characterised by a single scale, as discussed above. In the ordered region, the product of $2\to 3$ antennae is modulated by the correction factors $R_{2\to 4}$, to reproduce the full $2\to 4$ functions, and the two separate scales coincide as we approach the boundary, interpolating smoothly between the single-unresolved (iterated, strongly ordered) and double-unresolved (single-scale) limits.

\section{Explicit Construction of the 2$\to$4 Shower} \label{sect:2to4}

For a branching $1\ 2 \to  3\ 4\  5\  6$ we define the resolution scale as $Q_4 = 2 \min(p_\perp^{345},  p_\perp^{456})$, with $(p_\perp^{ijk})^2 = s_{ij}s_{jk}/s_{ijk}$. We let the direct $2\to 4$ shower populate all configurations for which the clustering corresponding to $Q_4$ is unordered. (Conversely, iterated $2\to 3$ splittings populate those configurations for which the clustering corresponding to $Q_4$ is ordered, with the correction factor $R_{2\to 4}$ reducing to $R_{2\to 4} \to a_4 / (a_3 a_3')$ when there is only a single ordered path, and, for gluon neighbours, the neighbour with the smaller resolution scale used to define $a_4$.)

We partition the direct $2\to 4$ phase space into two sectors: sector A with condition $p_\perp^{345} < p_\perp^{456}$ and sector B with $p_\perp^{345} > p_\perp^{456}$. 
For each sector, branching scales for $2\to 4$ emissions are generated from a uniformly distributed random number $R\in [0,1]$ by solving the following equation for $Q^2$: 
\begin{align}
    \label{eq:QE2eq}
    R = \Delta_{2\to4}(Q_{0}^2,Q^2)=\exp [-\mathcal{A}(Q_{0}^2,Q^2)]~.
\end{align}
This is done by means of the veto algorithm, which allows us to replace the (complicated) $a_4$ by a simple overestimate of it. We construct an appropriate ``trial function'' from the product of two eikonal functions $a_\mathrm{trial}^{2\to 3} = 2 g_s^2 C_A/p_\perp^2$, with an improvement factor $P_{\rm imp}$ from smooth-ordering showers~\cite{Giele:2011cb} which improves the approximation in the unordered region of phase space, and an overall factor 2 ensuring that the overestimate remains valid in the region $p_\perp^{345} \sim p_\perp^{456}$. Thus, 
\begin{align}
    \frac{1}{(16\pi^2)^2}a^{2\to4}_{\text{trial}} =&\ \frac{2}{(16\pi^2)^2}
    a_{\text{trial}}^{2\to 3}(Q_3^2)P_{\rm imp} a_{\text{trial}}^{2\to 3}(Q_4^2)
    \nonumber\\
    =&\ 
    \mathcal{C} \ \left(\frac{\alpha_s}{4\pi}\right)^2\frac{128} {(Q_{3}^2 + Q_{4}^2) Q_{4}^2}~.
\end{align}
where $ \mathcal{C}$ is the colour factor for the double branching, normalised so that $\mathcal{C} \to C_A^2$ at leading colour. In particular, the trial function for sector A~(B) is independent of momentum $p_6$~($p_3$) which makes it easy to translate the $2\to4$ phase spaces defined in \eqRef{eq:PS2to4} to shower variables. Technically, we generate these phase spaces by oversampling, vetoing configurations which do not fall in the appropriate sector.

For a fixed trial coupling $\hat{\alpha}_s$,   integration yields
\begin{multline}
    \mathcal{A}_{2\to4}^{\text{trial}}(Q_{0}^2,Q^2) =
     \mathcal{C}~I_\zeta \frac{\ln(2) \hat{\alpha}_s^2}{8\pi^2} \ln\frac{Q_{0}^2}{Q^2} \ln\frac{m^4}{Q_{0}^2 Q^2}~.
\end{multline}
where $I_\zeta$ is the $\zeta$ integral pertaining to the $4$-parton phase space, defined as for $p_\perp$ -ordering in ref.~\cite{Giele:2011cb}. The solution for $Q^2$ in eq.~(\ref{eq:QE2eq}) is thus 
\begin{align}
    Q^2 = 
    m^2 \exp\left(
    -\sqrt{\ln^2(Q_{0}^2/m^2) + 2f_R/\hat{\alpha}_s^2}
    \right)
\end{align}
where $f_R = -4\pi^2\ln R/(\ln(2)\mathcal{C}I_\zeta) $~. 

The trial generator can be made more efficient by including the leading effect of scaling violation, specifically the first-order running of $\alpha_s$,
\begin{align}
   \hat{\alpha}_s(k_\mu^2 p_\perp^2) = \frac{1}{b_0 \ln(k_\mu^2 p_\perp^2/\Lambda^2)}~,
\end{align}
where~$b_0 = (11 C_A - 4 n_f T_R)/(12\pi)$ and $k_\mu$ allows to apply a user-definable pre-factor. In the following equations we replace $k_\mu p_\perp$ by $k_\mu Q/2$.  The trial integral then becomes
\begin{multline}
    \mathcal{A}_{2\to4}^{\text{trial}}(Q_{0}^2,Q^2) =    \mathcal{C}~I_\zeta \frac{\ln(2)}{4\pi^2 b_0^2}
    \left[
    \frac{\ln(m^2/Q^2)}{\ln(k_\mu^2 Q^2/4\Lambda^2)}
    \right. \\ \left.
    -\frac{\ln(m^2/Q_{0}^2)}{\ln(k_\mu^2 Q_{0}^2/4\Lambda^2)}
    -\ln\left(
    \frac{\ln( k_\mu^2 Q_{0}^2/4\Lambda^2)}{\ln( k_\mu^2 Q^2/4\Lambda^2)}
    \right)
    \right]~.
\end{multline}
and the solution to eq.~(\ref{eq:QE2eq}) is 
\begin{align}
    Q^2 = \frac{4\Lambda^2}{k_\mu^2} \left(\frac{k_\mu^2 m^2}{4\Lambda^2} \right)^{-1/W_{-1}(-y)}~.
\end{align}
where 
\begin{align}
    y = \frac{\ln k_\mu^2 m^2/4\Lambda^2}{\ln k_\mu^2 Q_{0}^2/4\Lambda^2} \exp\left[-f_R b_0^2 -\frac{\ln k_\mu^2 m^2/4\Lambda^2}{\ln k_\mu^2 Q_{0}^2/4\Lambda^2}  \right]~,
\end{align}
and $W_{-1}(z)$ is the Lambert W function (solving $z=we^w$ for $w$ when $w\leq -1$)~for which we  use the numerical implementation of~\cite{lambertW1,lambertW2}. 

With a trial scale $Q$ having been generated, the remaining 4 kinematic variables (up to a global orientation) are generated according to the trial phase space integral in eq.~(\ref{eq:sud24}), allowing to construct explicit four-momenta. The sector veto is then applied and, if the sector is accepted, the trial is accepted with a probability 
\begin{align}
  P^{2\to4}_{\rm trial} = \frac{\alpha_s^{2}}{\hat{\alpha}_{s}^{ 2}}\frac{a_4}{a^{2\to4}_{\rm trial}}~,
\end{align}
where higher-order running effects can be included via the $\alpha_s$ ratio. Note that the final orientation of the post-branching system will also depend on the specific choice of kinematics map, see~\cite{Giele:2007di,GehrmannDeRidder:2011dm}.

The last piece required for the construction of the $2\to 4$ shower is the set of antenna functions, $a_4$, for $q\bar{q}$, $qg$, and $gg$ parent antennae. These are defined in the following section.

\section{Antenna Functions} \label{sect:antenna}

For a branching  $1~2 \to  3~4~5~6$ we consider partons $1$ and $2$ ($3$ and $6$) as the hard radiators (recoilers) and partons $4$ and $5$ as the radiated soft and/or collinear partons. (This is equivalent to the treatment of $2\to 3$ branchings.) These partons are colour-ordered and hence the antenna function for $3~4~5~6$ is not identical to that for  $3~5~4~6$. This is referred to as sub-antenna functions in the antenna-subtraction literature~\cite{GehrmannDeRidder:2005cm}. Since the shower framework is probabilistic, we also require that the antenna functions should be positive definite\footnote{We note that negative ones could in principle be treated using the formalism presented in~\cite{Platzer:2011dq,Hoeche:2011fd,Lonnblad:2012hz}.} (and bounded by the trial functions). For a $q\bar{q}$ parent antenna, the sub-antenna functions are equal to the full ones and we use $a_4^0$ from~\cite{GehrmannDeRidder:2005cm}.

For $qg$ and $gg$ parent antennae, the full leading-colour antenna functions in~\cite{GehrmannDeRidder:2005cm} contain several sub-antenna configurations corresponding to any quark-gluon or gluon-gluon pair as hard partons. Moreover, some of them  include terms representing two colour-unconnected emissions, for which the definition of the hard radiators and recoilers is ambiguous. The general problem of partitioning the full antenna functions into sub-antennae for colour-connected and colour-unconnected double emissions, with all singularities correct, is  nontrivial. Therefore, rather than using the full antenna functions,  we construct new sub-antennae based on $a_4^0$ by applying explicit modification factors to it, so that the unresolved limits agree with those of the relevant full antenna functions.

Specifically, for a $qg$ parent antenna, we define the sub-antenna $d_4^0$:
\begin{align}
   d_4^0 = \left\{ 
   \begin{array}{c l}
     a_4^0(3,4,5,6) \frac{d_3^0(\widehat{34},\widehat{45},6)}{a_3^0(\widehat{34},\widehat{45},6)} & \text{~Sec. A} \\
     a_4^0(3,4,5,6) \frac{d_3^0(3,\widehat{45},\widehat{56})}{a_3^0(3,\widehat{45},\widehat{56})} \frac{f_3^0(4,5,6)}{d_3^0(4,5,6)} & \text{~Sec. B} 
   \end{array}
   \right.,
\end{align}
where $a_3^0$, $d_3^0$, and $f_3^0$ are the single-emission $q\bar{q}$ and $qg$ (sub-)antenna functions defined in~\cite{GehrmannDeRidder:2005cm}, truncated so that only their singular terms are kept.
For a $gg$ parent antenna, we define the double-emission sub-antenna function as
\begin{align}
   f_4^0 = \left\{ 
   \begin{array}{c l}
     a_4^0(3,4,5,6) \frac{f_3^0(\widehat{34},\widehat{45},6)}{a_3^0(\widehat{34},\widehat{45},6)} \frac{f_3^0(3,4,5)}{d_3^0(3,4,5)} & \text{~Sec. A} \\
     a_4^0(3,4,5,6) \frac{f_3^0(3,\widehat{45},\widehat{56})}{a_3^0(3,\widehat{45},\widehat{56})} \frac{f_3^0(4,5,6)}{d_3^0(4,5,6)} & \text{~Sec. B} 
   \end{array}
   \right..
\end{align} 

In all but the triple-collinear limits, it can be shown analytically that our constructions of the sub-antenna functions, $d_4^0$ and $f_4^0$, exhibit the correct infrared singularities. In the limit in which three final-state partons are collinear, we have compared numerically with matrix elements and find good agreement.

Finally, we note that the original parton pair, $1\ 2$, is assumed to be on shell in the antenna formalism. For strongly ordered branchings, this is an excellent approximation, but for high-scale branchings, it was found in~\cite{Giele:2011cb} that by including an effective off-shellness term for the original parton pair, products of $2\to 3$ branchings could be brought into much better agreement with the full $2\to 4$ ones. Analogously, we may define an effective $2\to 5$ improvement factor, $Q_{2}^2/(Q_{2}^2+Q_4^2)$, which can be applied to the definitions of the 4-parton antenna functions whenever they appear in the context of 5- or higher-parton configurations, with $Q_2$ defined as the $3\to 2$ clustering scale of the parent partons together with their nearest colour neighbour.

 As a further numerical validation of our 4-parton antenna functions, away from the singular limits, we compare the leading-colour matrix element squared for $Z\to q_1 g_2g_3g_4\bar{q}_5$ with the approximation obtained using our sub-antenna function $d_4^0$, by defining the quantity
\begin{align}
\label{eq:r5}
   R_5 &= \frac{|M(Z\to q\bar{q})|^2}{ |M(1,2,3,4,5)|^2}\times
      \nonumber \\ & 
   \bigg(a_3^0(\widehat{123},\widehat{234},5) d_4^0(1,2,3,4) 
    \nonumber \\ & 
   +a_3^0(1,\widehat{234},\widehat{345}) d_4^0(5,4,3,2)
       \nonumber \\ & 
   +a_3^0(\widehat{12},\widehat{234},\widehat{45})d_3^0(1,2,\widehat{34}) d_3^0(5,4,3) 
       \nonumber \\ &  
   +a_3^0(\widehat{12},\widehat{234},\widehat{45})d_3^0(5,4,\widehat{23}) d_3^0(1,2,3) 
   \bigg)~,
\end{align} 
where the second and third lines represent the two possible $2\to 4$ branchings~($qg\bar{q}\to qggg\bar{q}$). Note that, in the sub-antenna functions $d_4^0$ only one sector, $A$ or $B$, contributes at a time, with the phase-space factorisations given by \eqRef{eq:PS2to4}. 
The fourth and fifth lines correspond to two colour-unconnected $2\to3$ branchings, correlations between which could only be taken into account properly at the $2\to 5$ level. 
For a $gg$ parent antenna, we compare the leading-colour matrix element for $H\to g_1 g_2 g_3 g_4$ to the result using our $f_4^0$ sub-antenna function, via the quantity 
\begin{align}
\label{eq:r4}
   R_4 =& \frac{|M(h\to gg)|^2}{ |M(1,2,3,4)|^2}
            \nonumber \\ & \hspace{-1cm}
   \times \left(
   f_4^0(1,2,3,4)+f_4^0(2,3,4,1)
      \right. \nonumber \\ & \hspace{-1cm} \left.  
  +f_4^0(3,4,1,2)+f_4^0(4,1,2,3) 
         \right. \nonumber \\ & \hspace{-1cm} \left.
         +f_3^0(\widehat{23},1,\widehat{34})f_3^0(2,3,4)
   +f_3^0(\widehat{34},2,\widehat{41})f_3^0(3,4,1) 
            \right. \nonumber \\ & \hspace{-1cm} \left.  
               +f_3^0(\widehat{41},3,\widehat{12})f_3^0(4,1,2)
   +f_3^0(\widehat{12},4,\widehat{23})f_3^0(1,2,3) 
   \right)~.
\end{align} 
Because the matrix element squared is symmetric under cyclic interchanges of the momenta, there are cases for which there is no way to define parent partons as radiators for $2\to4$ branchings, represented  by the fourth and fifth lines in eq.~(\ref{eq:r4}). 

\begin{figure*}[ht]
  \centering
  \includegraphics[scale=0.35]{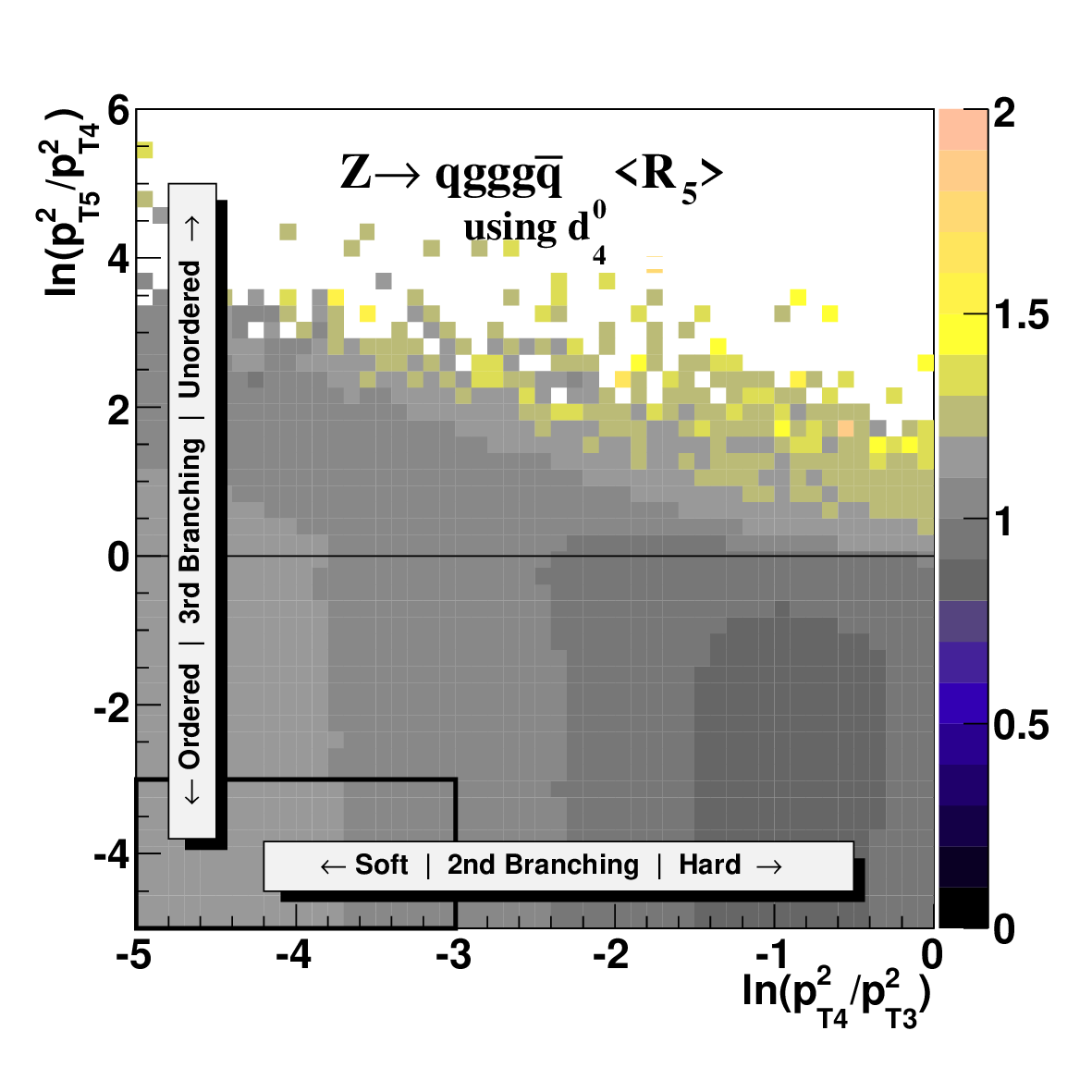} 
  \includegraphics[scale=0.35]{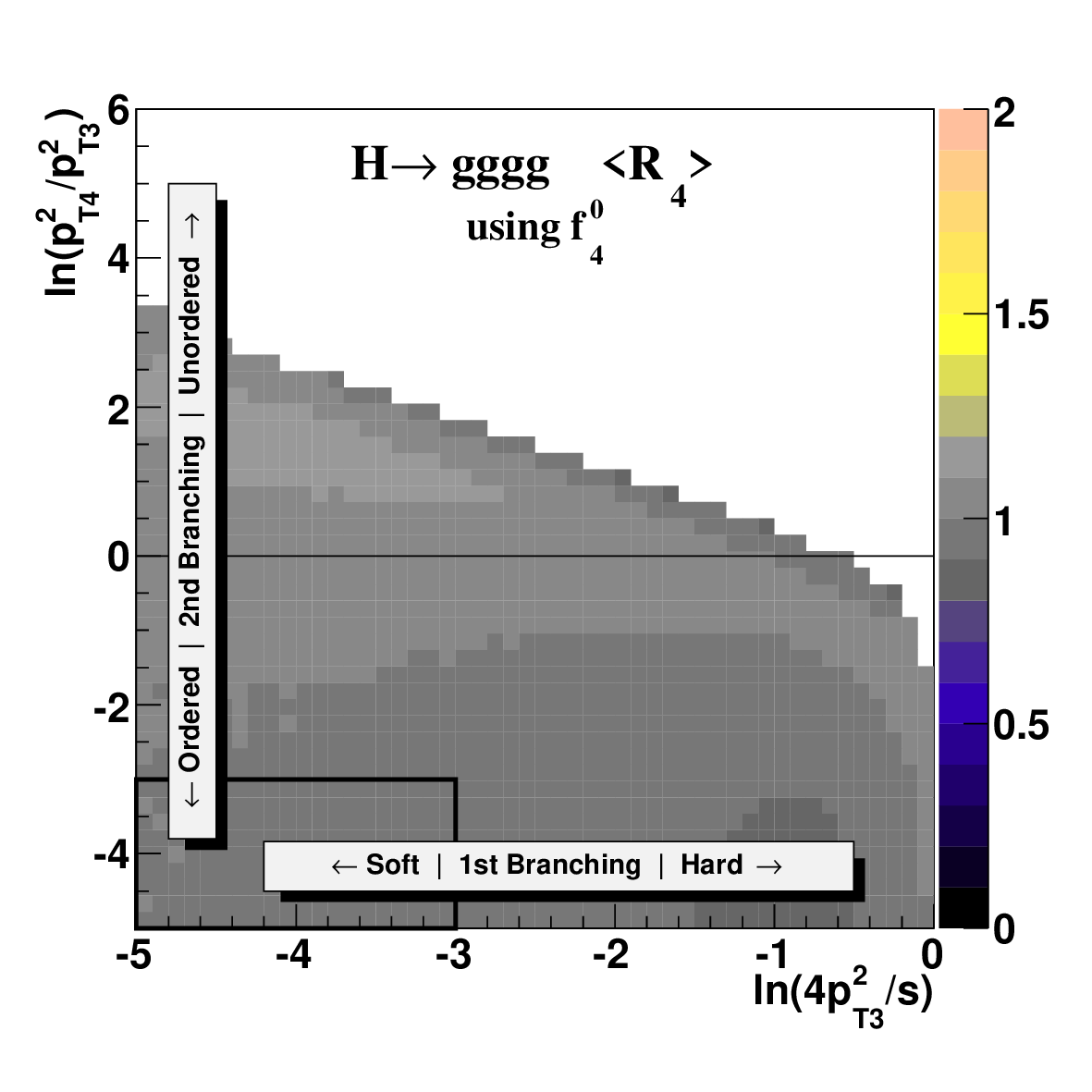}  
  \vspace{-2ex}
  \caption{\label{fig:d40f40}
  Mean value of $R_5$ (left) and $R_4$ (right) differentially over phase space, with $n$-parton clustering scales $p_{Tn} \equiv p_{T}(\Phi_n)$. Note that the ``top edges'' of the phase spaces correspond to hard configurations that are not logarithmically enhanced. 
  }
\end{figure*}
Results for $R_5$ and $R_4$ are shown in the left- and right-hand panes of fig.~\ref{fig:d40f40}, respectively. 
We use a smoothly ordered $2\to 3$ shower to generate phase space, and show the average values of $R_i$, differentially in the resolution scales for the last two branchings (averaging over the other phase-space variables of 5- and 4-parton phase space, respectively). As can be seen the effective sub-antennae $d_4^0$ and $f_4^0$ agree well with the matrix elements in both the ordered and unordered regions of phase space. Note that, at the very ``top'' of phase-space there is no overall scale hierarchy and hence no logarithmic enhancements. (The result in this region depends on a non-singular term which could be fixed by matching to the relevant fixed-order matrix elements.)

\section{Numerical Results} \label{sect:numeric}

\begin{figure*}[ht]
  \centering
  \includegraphics[scale=0.35]{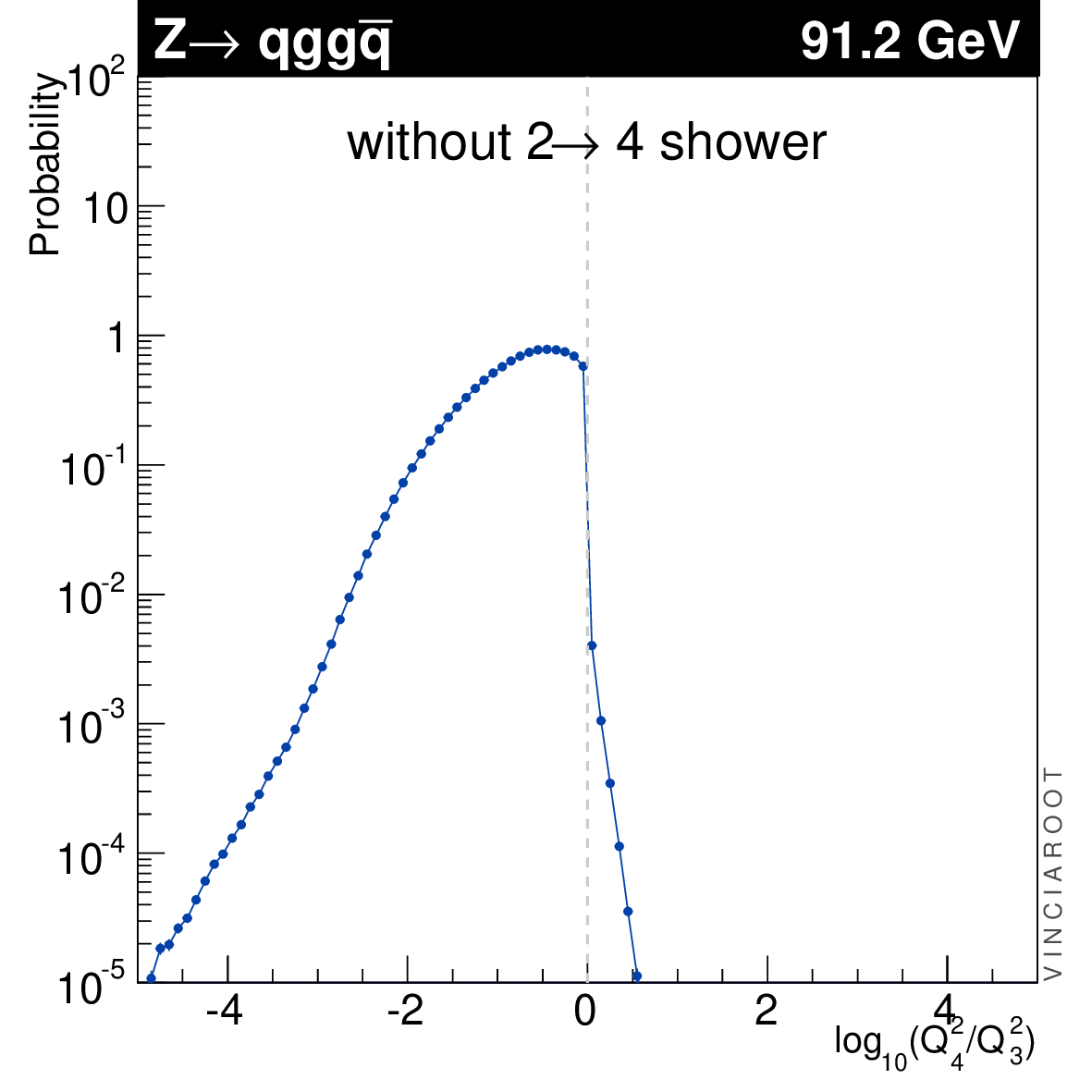} 
  \includegraphics[scale=0.35]{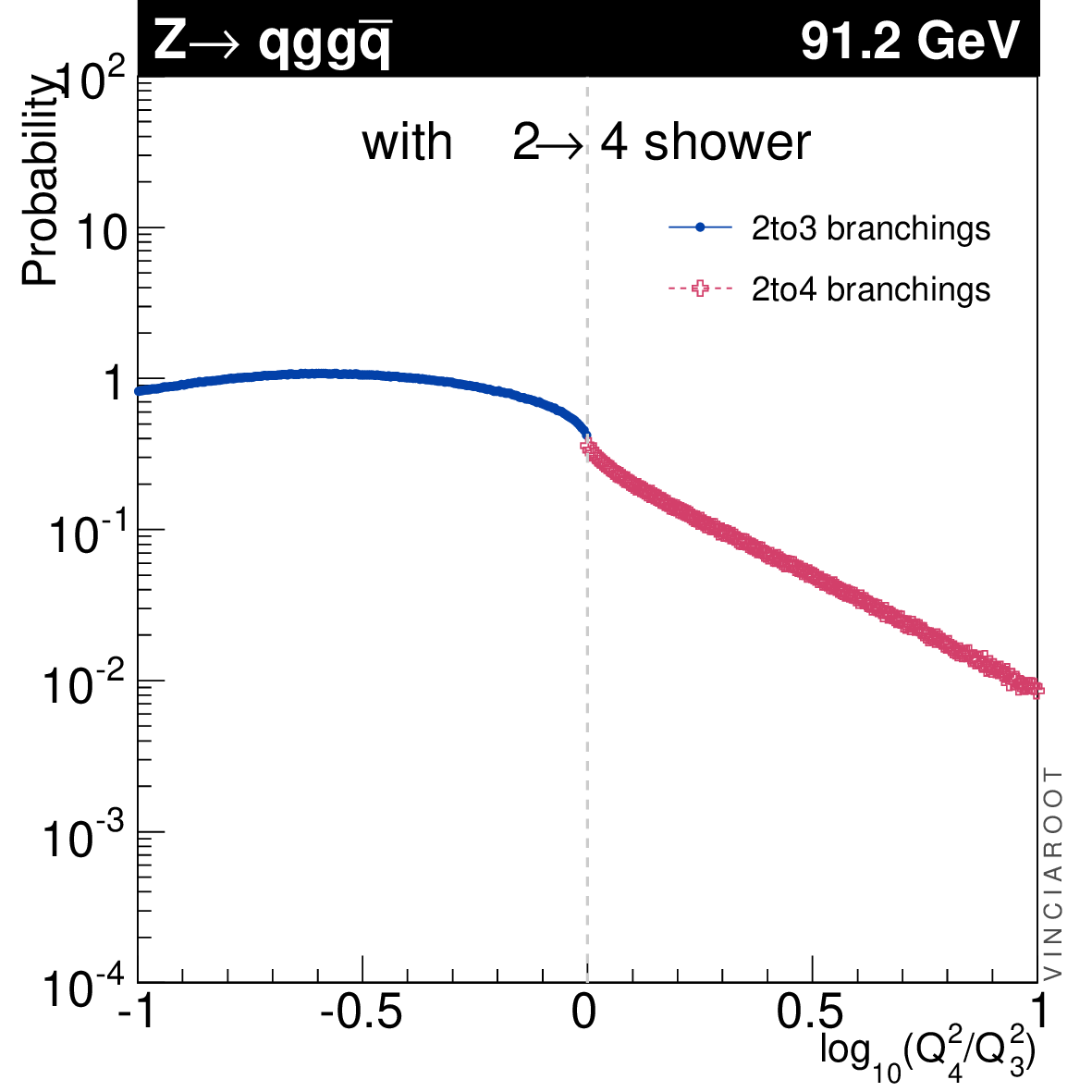} \\
  \includegraphics[scale=0.35]{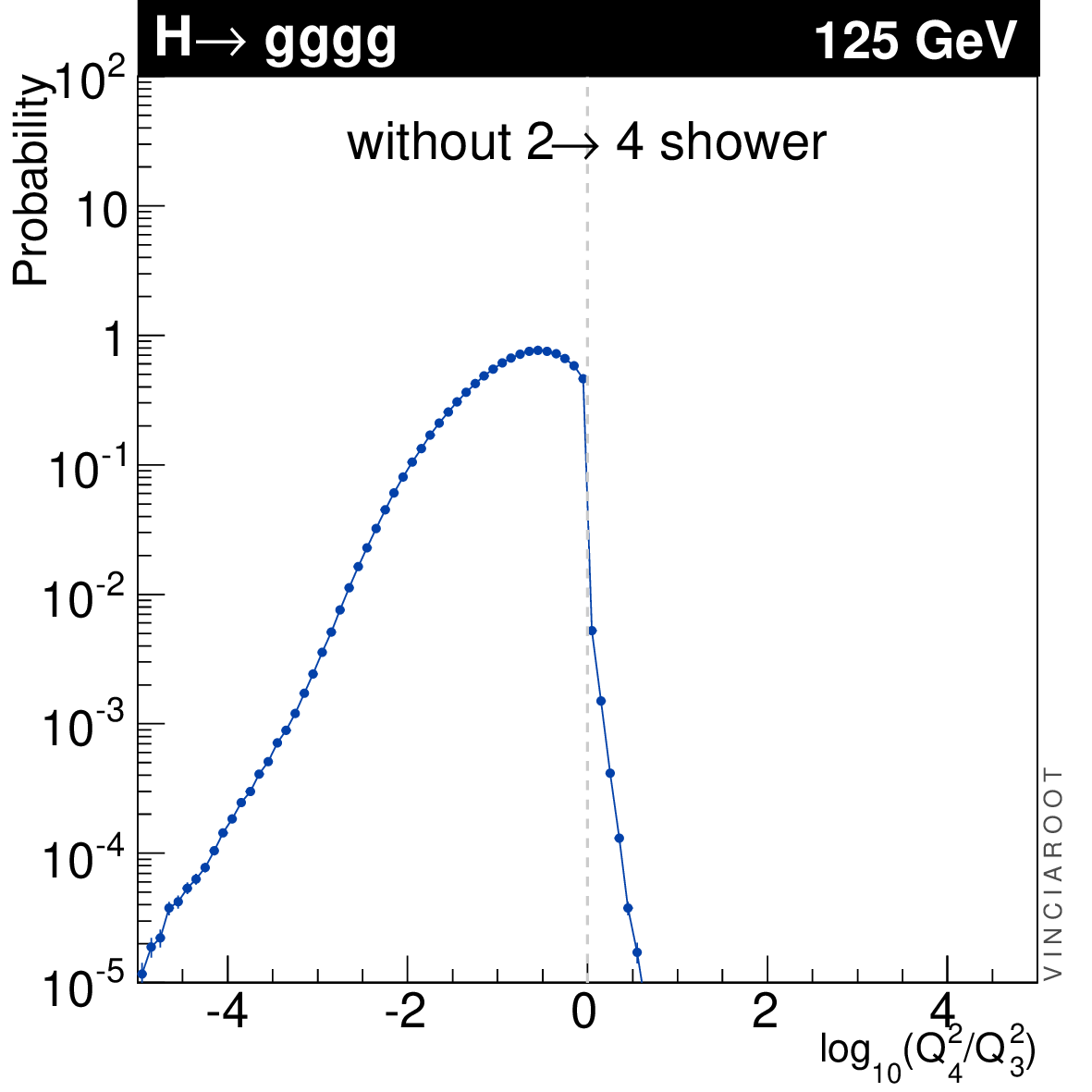} 
  \includegraphics[scale=0.35]{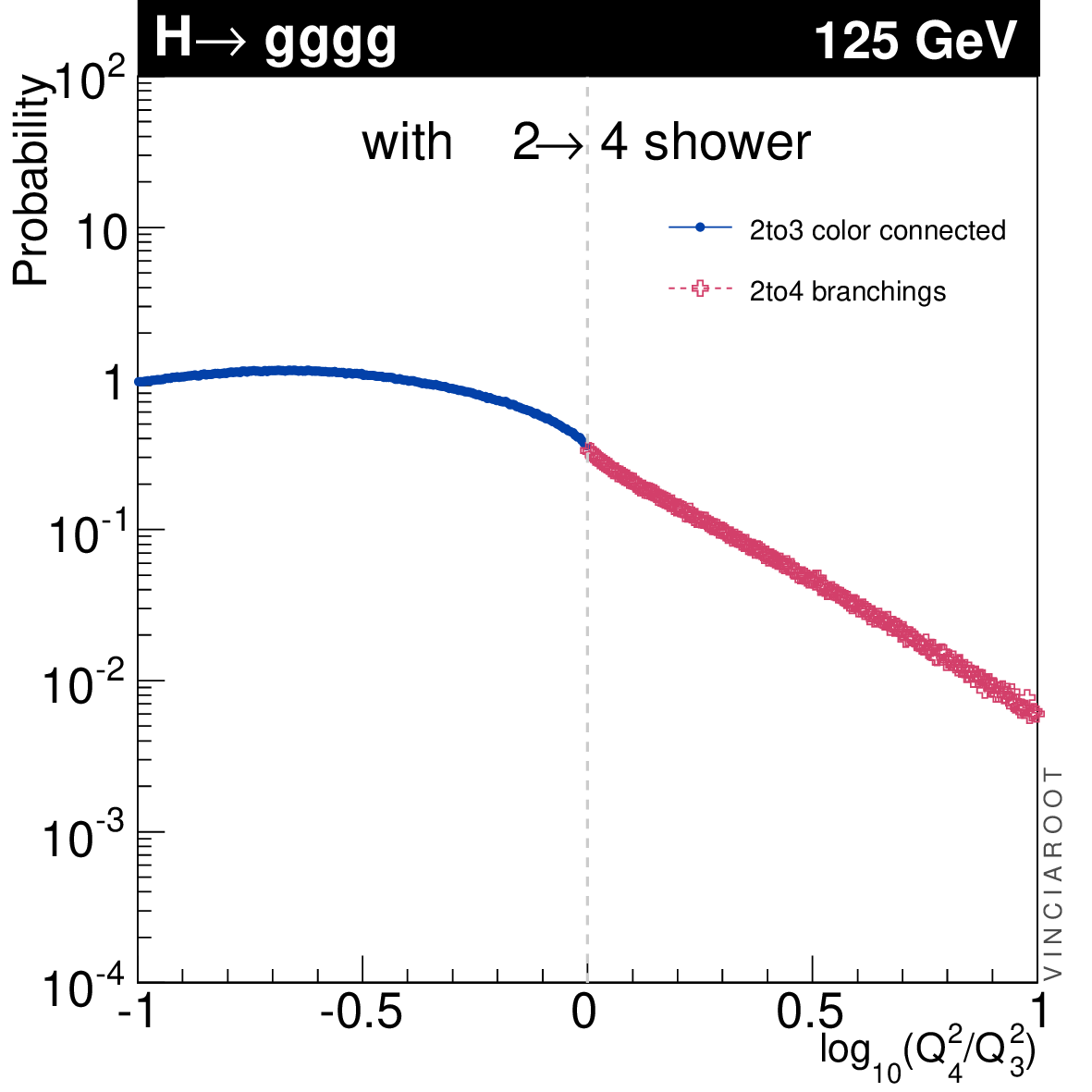} 
  \vspace{-2ex}\\
  \caption{\label{fig:zdecay} Top left: the ratio of sequential clustering scales $Q_4/Q_3$ for a strongly ordered $2\to 3$ shower, for $Z\to q g g \bar{q}$ (on log-log axes). Top right: closeup of the region around $Q_4/Q_3 \sim 1$, with 2$\to$4 branchings included. Bottom row: the same for $H\to gggg$. \label{fig:hdecay}.
  }
\end{figure*}

We restrict ourselves to strong ordering in the resolution scale, defined as the scale of the last radiation. That is, a $2\to 4$ branching can have an ``unordered'' intermediate step, but the resolution scale of the resulting 4-parton configuration will still be required to be lower than that at which the parent 2-parton antenna was created. Notice that, due to momentum recoil effects, the local definition of the evolution scale (inside a branching antenna) might not be the smallest global scale in a given event. (In general, the local and global resolution scale is only guaranteed to agree for so-called sector showers~\cite{LopezVillarejo:2011ap}.)

The number of direct $2\to4$ branchings is suppressed by the small volume of unordered phase space. For a $q\bar{q}$ parent antenna, the probability of a direct $2\to4$ branching is around 5$\%$~. The effect of the $2\to 4$ shower on inclusive distributions is therefore small. Since we moreover only have partial $\mathcal{O}(\alpha_s)$ corrections in the Sudakov factor, we focus on theory-level distributions to further validate the $2\to 4$ implementation, postponing a more detailed comparison with physical observables to future work.

To compare $2\to3$ and $2\to4$ showers, we truncate the shower such that every event has at most four partons in the final state and we select the events with two gluon emissions for original colour dipole. Using shower history and colour information we identify the two radiated gluons and define the smaller transverse momentum associated with these two gluons emissions as $Q_4/2$. For example in branching $1~2\to 3~4~5~6$~ where $4$ and $5$ are radiations $Q_4=2\min(p_\perp^{345}, p_\perp^{456})$~. For the case of $4$ and $6$ as emissions which corresponds to two iterated colour-unconnected $2\to3$ branchings,  $Q_4=2\min(p_\perp^{345}, p_\perp^{365})$~. The intermediate states are obtained by clustering three partons to two intermediate ones according to the choice of $Q_4$. And $Q_3$ is the scale of the other emitted parton, in the intermediate (3-parton) state. With this definition, all direct $2\to 4$ branchings satisfy the condition $Q_4>Q_3$~ and most $2\to3$ branchings satisfy the complementary condition $Q_4<Q_3$. In order to make the theory prediction consistent on the boundary $Q_4=Q_3$, we modify the last $2\to 3$ branching by using the strong coupling at the scale of $Q_4$, thus making the product $\alpha_s^2(Q_4)$ the same for branchings on either side of the ordering threshold.

Figure~\ref{fig:zdecay} shows the distribution of the ratio of $Q_4/Q_3$ for $Z$ decaying into $ q g g \bar{q}$ (top row) and Higgs decaying into $gggg$ (bottom row). It can be seen that, due to momentum recoil effects, the pure $2\to3$ shower \emph{can} generate some (highly suppressed) contributions in the phase-space region $Q_4>Q_3$. In the $2\to 4$ shower, the iterated $2\to3$ branchings are matched by sub-antenna functions $a_4^0$ or $f_4^0$, as discussed above, and the direct $2\to 4$ branchers are used to populate the unordered phase space. For $Z$ boson decay, the iterated $2\to3$ and $2\to 4$ branchings are effectively generated by the same function $a_4^0$. However, for Higgs boson decay, the $2\to 4$ functions only include colour-connected double emissions, while the presence of two antennae in the Born configuration means that the iterated $2\to 3$ branchings can also generate two colour-unconnected emissions, which are not matched to $2\to 4$. In order to clarify the correctness of the $2\to 4$ implementation in fig.~\ref{fig:zdecay} we do not include the contribution from colour-unconnected branching sequences in the $2\to 3$ contributions for the case with $2\to 4$ shower. As shown in the right-hand plots, the $2\to 4$ shower fills in the unordered phase space, and, in the limit $Q_4 \sim Q_3$, consistently matches onto the $2\to3$ result.

\section{Conclusion and Outlook} \label{sect:conclu}
We have presented a framework for deriving corrections at the NLO level to Sudakov form-factor integrands, which generates $2\to 3$ and $2\to 4$ strongly-ordered showers. Compared to matching and merging methods for each branching, our corrections are  generated directly by the Sudakov form factor and are present throughout the shower evolution. We hope that this framework may serve as a useful conceptual step towards the resummation of further (subleading) logarithmic terms in parton showers.

A proof-of-concept implementation of NLO corrections to a single gluon emission from a $q\bar{q}$ antenna was presented in \cite{Hartgring:2013jma}. A crucial new ingredient developed here for the first time are direct $2\to 4$ branchings, for which we have presented an explicit Sudakov-type phase-space generator, in which the resolution scale of the 4-parton state is used as the shower evolution measure. We applied this to the case of a colour antenna radiating two (non-strongly-ordered) gluons, via a decomposition of the phase space into two sectors. For each sector we construct a trial function and trial integral based on iterated $2\to 3$ ones, with the scale of the intermediate 3-parton state integrated over. (I.e., the intermediate 3-parton resolution scale is only used to separate what is ordered --- and hence accessible by the iterated $2\to3$ evolution --- from what is unordered.) 
We also define sub-antenna functions for dipole-antennae in which one or both of the parent partons are gluons, starting from the antenna function for quark-antiquark pairs, which is a good first approximation to the amplitude squared. As a validation, we compare $2\to 4$ and $2\to 3$ branchings in~fig.~\ref{fig:zdecay}.  As expected, the $2\to 4$ branchings extend the phase-space population into the unordered region. Importantly, the $2\to 4$ and $2\to 3$ branchings produce consistent results on the boundary $Q_4=Q_3$.

In the near future we will extend the $2\to 4$ shower formalism to include $g\to q\bar{q}$ splittings. We also expect to include the second-order correction to the $2\to 3$ Sudakov form factor defined in eq.~(\ref{eq:2to3sudakov}). Finally, in the longer term we plan to turn our attention to the initial state,  extending the formalism to the case of hadron collisions.

\section*{Acknowledgements} We are grateful to B.~El-Menoufi for
spotting that the term $\Delta^1$ had been omitted in the definitions of
eqs.~(\ref{eq:sudNLO}) and (\ref{eq:nlo_su}) in v1 and v2 of this
manuscript. Fortunately, eq.~(\ref{eq:2to3sudakov}) was
still correct and hence the subsequent derivations and results were
not changed. We are also grateful to S.~Prestel for comments on the
original manuscript. 
This work was supported in part by
the ARC Centre of Excellence for Particle Physics at the Terascale
(CoEPP). PS is the recipient of an ARC Future Fellowship, FT130100744. 

%\appendix

\bibliographystyle{JHEP}

\bibliography{letter2to4}

\end{document}